%% file: main.tex
\documentclass[sigplan,dvipsnames]{acmart}
\acmConference[ABC '00]{A Boilerplate Conference}{January 0000}{}
\setcopyright{none}
\usepackage{graphicx}
\usepackage{booktabs}
\usepackage{subcaption}
\usepackage{mathtools}
\usepackage{dsfont}
\usepackage{paralist}
\usepackage{caption}
\usepackage{amscd}
\usepackage{tikz}
\usetikzlibrary{calc,tikzmark,arrows.meta,automata,positioning,shadows,decorations.text,shapes.geometric,shapes.multipart,chains}
\usepackage{multirow}
\usepackage{dashbox}
\usepackage{amsfonts}
\usepackage{fontawesome}
\usepackage{appendix}
\usepackage{xspace}
\usepackage{algorithm}
\usepackage{algorithmicx}
\usepackage{multicol}
\usepackage[noend]{algpseudocode}
\MakeRobust{\Call}
\algrenewcommand\algorithmicrequire{\textbf{Input:}}
\algrenewcommand\algorithmicensure{\textbf{Output:}}
\usepackage{amsmath} %for the \xrightsquigarrow
\usepackage[mathb]{mathabx}
\usepackage{hyperref}
\usepackage{listings}
\usepackage[utf8x]{inputenc}
\usepackage[T1]{fontenc}
\DeclareUnicodeCharacter{1EC5}{\~{\^e}}
\usepackage{stmaryrd} % llparenthesis
\usepackage{stackrel}
\usepackage{todonotes}
\usepackage[forwardlinking=no]{apxproof}

\usepackage{version}
\usepackage{cancel} % slash over symbol
\usepackage[capitalise]{cleveref}
\newcommand{\crefpart}[2]{%
  \hyperref[#2]{\namecref{#1}~\labelcref*{#1}~\ref*{#2}}%
}
\usepackage[inline,shortlabels]{enumitem}
\usepackage{framed}

\usepackage[commandnameprefix=ifneeded]{changes}
\usepackage{tabularx}
\usepackage{array}
\newcolumntype{L}{>{$}l<{$}} % math-mode version of "l" column type

\usepackage{mathpartir}

\def\OPTIONConf{1}
\usepackage{jdunfield}
\usepackage{xifthen}
\usepackage{amsthm}
\theoremstyle{definition}
\newtheoremrep{theorem}{Theorem}[section]
\newtheoremrep{lemma}[theorem]{Lemma}
\newtheoremrep{definition}[theorem]{Definition}
\newtheoremrep{corollary}[theorem]{Corollary}
\newtheoremrep{remark}[theorem]{Remark}
\newtheoremrep{conjecture}[theorem]{Conjecture}
\newtheoremrep{example}[theorem]{\mbox{\emph{Example}}}

\definecolor{kwd}{HTML}{3F6FA8}
\definecolor{ndkwd}{HTML}{E69F00}
\colorlet{err}{Red}
\colorlet{warn}{Dandelion}

\usepackage{dafny}
\input{macros}
\begin{document}

%% Title information
%\title[Short Title]{}         %% [Short Title] is optional;
                                        %% when present, will be used in
                                        %% header instead of Full Title.
\title{From Traces to Program Incorrectness: A Type-Theoretic Approach}
%\titlenote{with title note}             %% \titlenote is optional;
                                        %% can be repeated if necessary;
                                        %% contents suppressed with 'anonymous'
%\subtitle{Subtitle}                     %% \subtitle is optional
%\subtitlenote{with subtitle note}       %% \subtitlenote is optional;
                                        %% can be repeated if necessary;
                                        %% contents suppressed with 'anonymous'

%%
%% The "author" command and its associated commands are used to define
%% the authors and their affiliations.
%% Of note is the shared affiliation of the first two authors, and the
%% "authornote" and "authornotemark" commands
%% used to denote shared contribution to the research.

\author{Yongwei Yuan}
%\authornote{with author1 note}          %% \authornote is optional;
                                        %% can be repeated if necessary
\orcid{0000-0002-2619-2288}             %% \orcid is optional
\affiliation{
  \institution{Purdue University}            %% \institution is required
  \country{USA}                    %% \country is recommended
}
\email{yuan311@purdue.edu}          %% \email is recommended

\author{Zhe Zhou}
\orcid{0000-0003-3900-7501}
\affiliation{
  \institution{Purdue University}            %% \institution is required
  \country{USA}                    %% \country is recommended
}
\email{zhou956@purdue.edu}          %% \email is recommended

\author{Julia Belyakova}
\orcid{0000-0002-7490-8500}
\affiliation{
  \institution{Purdue University}            %% \institution is required
  \country{USA}                    %% \country is recommended
}
\email{julbinb@gmail.com}

\author{Benjamin Delaware}
\orcid{0000-0002-1016-6261}
\affiliation{
  \institution{Purdue University}            %% \institution is required
  \country{USA}                    %% \country is recommended
}
\email{bendy@purdue.edu}

\author{Suresh Jagannathan}
\orcid{0000-0001-6871-2424}
\affiliation{
  \institution{Purdue University}            %% \institution is required
  \country{USA}                    %% \country is recommended
}
\email{suresh@cs.purdue.edu}
%
%% By default, the full list of authors will be used in the page
%% headers. Often, this list is too long, and will overlap
%% other information printed in the page headers. This command allows
%% the author to define a more concise list
%% of authors' names for this purpose.
%\renewcommand{\shortauthors}{Trovato and Tobin, et al.}

%% Abstract
%% Note: \begin{abstract}...\end{abstract} environment must come
%% before \maketitle command
\begin{abstract}
\input{abstract}
\end{abstract}

%%% 2012 ACM Computing Classification System (CSS) concepts
%%% Generate at 'http://dl.acm.org/ccs/ccs.cfm'.
%\begin{CCSXML}
%<ccs2012>
%<concept>
%<concept_id>10011007.10011006.10011008</concept_id>
%<concept_desc>Software and its engineering~General programming languages</concept_desc>
%<concept_significance>500</concept_significance>
%</concept>
%<concept>
%<concept_id>10003456.10003457.10003521.10003525</concept_id>
%<concept_desc>Social and professional topics~History of programming languages</concept_desc>
%<concept_significance>300</concept_significance>
%</concept>
%</ccs2012>
%\end{CCSXML}
%
%\ccsdesc[500]{Software and its engineering~General programming languages}
%\ccsdesc[300]{Social and professional topics~History of programming languages}
%% End of generated code

%% Keywords
%% comma separated list
%\keywords{keyword1, keyword2, keyword3}  %% \keywords are mandatory in final camera-ready submission

%% \maketitle
%% Note: \maketitle command must come after title commands, author
%% commands, abstract environment, Computing Classification System
%% environment and commands, and keywords command.
\maketitle
\input{tpsa}

\bibliography{references,bibliography}
% If your work has an appendix, this is the place to put it.
\newpage

%% Appendix
\appendix

\end{document}

%% file: macros.tex
\newcommand{\kwd}[1]{{\color{kwd}\texttt{\bfseries #1}}}

\newcommand{\EM}[1]{\ensuremath{#1}\xspace}

\newcommand{\myidx}[1]{%
  \mathsf{#1}
}
\newcommand{\myidxwrapper}[2]{%
  \ifthenelse{\isempty{#2}}
  {\EM{#1}}
  {\EM{#1_{\myidx{#2}}}}%
}

\newcommand{\darr}{{\dasharrow}}

\newcommand{\eff}[1]{\texttt{#1}}

\newcommand{\metavar}[2]{\EM{#1_\mathsf{#2}}}

\let\oldtau\tau
\renewcommand{\tau}[1][]{\metavar{\oldtau}{#1}}

\newcommand{\letin}[2]{\EM{\kwd{let}\ #1 = #2\ \kwd{in}}\,}

\newcommand{\inangle}[1]{\EM{\langle #1 \rangle}}
\newcommand{\inparen}[1]{\left(#1\right)}

\newcommand{\tUnder}[1]{\EM{{\color{RoyalBlue} \left[ {\color{black}#1} \right]}}}

\newcommand{\triple}[3]{\EM{
    {[}#1{]} #2 {[}#3{]}
  }}

\newcommand{\type}[1]{\EM{\mathsf{#1}}}

\newcommand{\lneg}{{\EM{\backsim}}}

\newcommand{\deriv}[3][]{%
  \EM{\mathsf{d}_{#2}\ifx r#1\inparen{#3}\else #3\fi}%
}
\newcommand{\Deriv}[3][]{%
  \EM{\mathsf{D}_{#2}\ifx r#1\inparen{#3}\else#3\fi}%
}

\newcommand{\trace}[1][]{%
  \ifthenelse{\isempty{#1}}
  {\EM{\pi}}
  {\EM{\pi_{\textsf{#1}}}}\xspace
}
\newcommand{\Trace}[1][]{%
  \ifthenelse{\isempty{#1}}
  {\EM{\Pi}}
  {\EM{\Pi_{\textsf{#1}}}}\xspace
}
\let\oldPhi\Phi
\renewcommand{\Phi}[1][]{%
  \ifthenelse{\isempty{#1}}
  {\EM{\oldPhi}}
  {\EM{\oldPhi_{\textsf{#1}}}}\xspace
}
\let\oldphi\phi
\renewcommand{\phi}[1][]{%
  \ifthenelse{\isempty{#1}}
  {\EM{\oldphi}}
  {\EM{\oldphi_{\textsf{#1}}}}\xspace
}
\let\oldGamma\Gamma
\renewcommand{\Gamma}[1][]{\myidxwrapper{\oldGamma}{#1}}
\let\oldTheta\Theta
\renewcommand{\Theta}[1][]{\myidxwrapper{\oldTheta}{#1}}

\newcommand{\FA}[1][]{\myidxwrapper{\mathcal{A}}{#1}}

%% file: abstract.tex
We present a type-theoretic framework for reasoning about
incorrectness in functional programs that interact with effectful,
opaque library APIs.  Our approach centers on traces --
temporally-ordered sequences of library API invocations -- which
naturally characterize both the preconditions of individual APIs and
their composite behavior.  We represent these traces using symbolic
regular expressions (SREs), enabling formal specification of incorrect
abstract data type (ADT) behaviors across function boundaries.  The
core contribution is a novel type inference algorithm that operates
modulo specified incorrectness properties and leverages the symbolic
finite automata (SFAs) representations of regexes for compositional
reasoning of traces.  When the algorithm succeeds, the inferred types
witness that an ADT implementation can exhibit some subset of the
specified incorrect behaviors.  This represents the first systematic
approach to underapproximate reasoning against trace-based
incorrectness specifications, enabling a new form of trace-guided
compositional analysis.

% Recent trends in program analysis have shifted towards
% \emph{incorrectness reasoning}, which approaches static analysis
% through the lens of bug finding rather than proving the absence of
% bugs.  By summarizing underapproximate program behaviors, tools based
% on incorrectness principles achieve compositional analysis while
% eliminating false-positive bug reports.  However, existing bug-finding
% tools primarily focus on localized bugs associated with specific
% program points (\eg, potential null pointer dereferences) rather than
% temporal properties spanning multiple function scopes.

% In this work, we present a novel refinement type system where types
% serve as true witnesses to program incorrectness involving
% temporally-ordered events across function boundaries.  Specifically,
% we specify program incorrectness using symbolic finite automata (SFAs)
% that encapsulate behaviors of interest and accept traces of events
% manifesting these behaviors.  We develop a type inference algorithm
% that operates modulo an incorrectness SFA, with the inferred types
% witnessing that the program can exhibit a subset of the specified
% behaviors of interest.  Our approach is a first attempt at
% underapproximate reasoning against trace-based incorrectness
% specification and enables a novel form of trace-guided compositional
% reasoning.

%%% Local Variables:
%%% mode: LaTeX
%%% TeX-master: "main"
%%% End:

%% file: tpsa.tex
\section{Traces as Incorrectness Specifications}

Traces of events are a popular and effective formalism for modeling and reasoning about the behaviors of effectful functional programs
\cite{skalkaHistoryEffectsVerification2004,songAutomatedTemporalVerification2022,sekiyamaTemporalVerificationAnswerEffect2023,zhouHATTrickAutomatically2024}.
\emph{Global} safety properties of such programs can be captured as desirable sequences of temporally-ordered program events that cross function boundaries. While such specifications are expressive and amenable to a variety of reasoning techniques, formulating safety properties in terms of traces can involve non-trivial effort. 

% A $\eff{get}~k$ event, when succeeding a corresponding $\eff{put}~k~v$
% event followed by events that does not interfere with the entry
% associated with $k$, is expected to return the stored value $v$.  Our
% first attempt for specifying this safety property in regular
% expressions (regexes) is
% \begin{equation*}
%   \bullet^*\ \inangle{\eff{put}~k~v}\
%   (\lneg\inangle{\eff{put}~k~\_})^*\
%   \inangle{v{\gets}\eff{get}~k}\ \bullet^*
% \end{equation*}
% which states that all safe traces should be admissible by this regex
% for all instantiations of $k$ and $v$.  The regex, however,
% over-approximates the set of safe traces because it does not specify

Consider a linked-list implementation that maintains the ordering
relation among nodes in the list using an effectful key-value store
(named $\eff{Nxt}$).  The key-value store provides two APIs,
$\eff{put} : \type{addr}\arr\type{addr}\arr\type{unit}$ and
$\eff{get} : \type{addr}\arr\type{addr}$, where type \type{addr}
denotes the addresses of nodes.  A safety property concerning the
protocol between the linked-list and its underlying $\eff{Nxt}$ store
states that each node must have a unique parent at all time.  Such a
property can be expressed in terms of temporally-ordered sequences of
API invocations, namely \emph{traces}; any $\eff{Nxt.put}$ event that
links node $a$ to node $b$, denoted by $\inangle{\eff{Nxt.put}~a~b}$,
can be followed by sequences of events that does not link nodes other
than $a$ to $b$, denoted by
$(\lneg\inangle{\eff{Nxt.put}~\cancel{a}~b})^*$, or does not link
nodes other than $a$ to $b$ until $b$ is unlinked from $a$, denoted by
$\inangle{\eff{Nxt.put}~a~\cancel{b}}$.

One may attempt to regard this unique-parent safety property as a
representation invariant that holds at any program point
\cite{zhouHATTrickAutomatically2024} and represent it as a regular
expression (regex)
\begin{multline}\label{eq:over-safe-uniq}
  \bullet^*\ \inangle{\eff{put}~a~b}\ (
  (\lneg\inangle{\eff{put}~\cancel{a}~b})^*\ \vee \\
  \left((\lneg\inangle{\eff{put}~\cancel{a}~b})^*\
  \inangle{\eff{put}~a~\cancel{b}}\ \bullet^*\right)
  )
\end{multline}
stating that for all pairs of nodes $a$ and $b$, all traces accepted
by the regex (with the name of key-value stores omitted for
simplicity) is safe.  The statement, unfortunately, is not true, as
illustrated by an unsafe trace that \cref{eq:over-safe-uniq}
unintentionally accepts:
\begin{equation}\label{eq:bad-trace-leading}
  {\color{err}\inangle{\eff{put}~1~2}\ \inangle{\eff{put}~3~2}}\
  \inangle{\eff{put}~3~4}\ \inangle{\eff{put}~1~2}
\end{equation}
where we represent node addresses using intergers for ease of
discussion.  It is obvious that node $1$ and $3$ can both be the
parent of node $2$ at the same time within
\cref{eq:bad-trace-leading}, violating the safety property.  Checking
that \cref{eq:bad-trace-leading} is eliminated from a hand-written
safety specification like \cref{eq:over-safe-uniq}, on the other hand,
requires us to consider
\begin{enumerate*}[(i)]
\item all pairs of $a$ and $b$ and
\item all sorts of events that may precede the
  $\inangle{\eff{put}~a~b}$ of our interest as the leading $\bullet^*$
  accepts all prefix traces.
\end{enumerate*}
Among other choices, only when $a$ is $1$, $b$ is $2$, and the last
$\inangle{\eff{put}~1~2}$ matches $\inangle{\eff{put}~a~b}$, does
\cref{eq:over-safe-uniq} accept \cref{eq:bad-trace-leading}.  Jumping
through such hoops of \emph{universal} reasoning and coming up with a
regex that \emph{precisely} captures the safety property deems
difficult.

One resort to the universal reasoning problem is a manual
compositional approach where user provide a pair of regexes for each
method under analysis, denoting the required context under which the
method can be invoked and the allowed effect the method can produce
respectively \cite{yuanDerivativeGuidedSymbolicExecution2024}.  For a
method that manipulates an existing linked list, say removing a node
with a given payload from the list, it is reasonable to consider any
existing pair of linked nodes $a$ and $b$ prior to the method
invocation, as specified by \newcommand{\rgxLinkTo}[2]{\EM{\FA_{#1
      \curvearrowright #2}}}
\newcommand{\refLinkTo}[2]{\hyperref[link-a-to-b]{\rgxLinkTo{#1}{#2}}}
\begin{equation}\label{link-a-to-b}
  \bullet^*\ \inangle{\eff{put}~a~b}\ (\lneg\inangle{\eff{put}~a~\_})^*
  \tag{$\rgxLinkTo{a}{b}$}
\end{equation}
and check that the set of traces admissible by
\begin{equation*}
  (\lneg\inangle{\eff{put}~\cancel{a}~b})^*\ \vee \\
  \left((\lneg\inangle{\eff{put}~\cancel{a}~b})^*\
  \inangle{\eff{put}~a~\cancel{b}}\ \bullet^*\right)
\end{equation*}
includes those that can be produced by the method.  The context regex
\refLinkTo{a}{b} helps fixate the linked pairs of nodes whose
unique-parent property is to be checked, alleviating the burden of
universal reasoning.  However, this decomposition not only requires
user to specialize the safety property to each method but also is
subject to changes stemmed from the safety property or the
implementation, which is cumbersome at best.

% Compositional reasoning of such a regex is nonexistent in an automatic
% verification tool like \Marple, as all verification conditions may
% involve the regex \cite{zhouHATTrickAutomatically2024}.  In the case
% that the regex underapproximates the intended safety property, \eg,
% the first disjunct is dropped, deriving counterexamples and refining
% the regex with them is thus a rather slow process.
% \begin{equation}\label{eq:bad-trace-trailing}
%   \inangle{\eff{put}~1~2}\ \inangle{\eff{put}~1~3}\
%   \inangle{\eff{put}~1~2}\ \inangle{\eff{put}~3~2}
% \end{equation}
\newcommand{\ofValue}[2]{\EM{\FA_{#1 : #2}}}

To address these challenges, we exploit the dual of the unique-parent
property and formulate its incorrectness specification, also as a
regex:
\newcommand{\rgxNotUniq}[2]{\EM{\FA_{#1 \curvearrowright #2 \curvearrowleft \cancel{#1}}}}
\newcommand{\refNotUniq}[2]{\hyperref[not-unique]{\rgxNotUniq{#1}{#2}}}
\begin{equation}\label{not-unique}
  \bullet^*\ \inangle{\eff{put}~a~b}\
  (\lneg\inangle{\eff{put}~a~\_})^*
  \inangle{\eff{put}~\cancel{a}~b}\ \bullet^*
  \tag{$\rgxNotUniq{a}{b}$}
\end{equation}
This regex serves as an abstraction pattern that captures a family of
traces witnessing violations of the safety property and identifies a
node being linked to at least two parents.  In particular, the pattern
abstract over:
\begin{enumerate*}[(i)]
\item node addresses, where $a$ and $b$ represent arbitrary nodes
  linked via $\inangle{\eff{put}~a~b}$,
\item temporal context, where $\bullet^*$ represents arbitrary events before and after the violation, and
\item intermediate operations, where $(\lneg\inangle{\eff{put}~a~\_})^*$ represents any sequence of operations that do no intervene with the link from $a$ to $b$.
\end{enumerate*}
Take \cref{eq:bad-trace-leading} as an example, its first two events
(highlighted in red) match the incorrectness pattern --
$\inangle{\eff{put}~a~b}$ followed by
$\inangle{\eff{put}~\cancel{a}~b}$, with $1$, $2$, and $3$
instantiating $a$, $b$, and $\cancel{a}$ respectively.  This
exemplifies our \emph{existential} reasoning principle: for any unsafe
behavior, the existence of corresponding traces matching the
abstraction pattern suffices to demonstrate that the behavior is
accounted for.  By shifting from universal reasoning over safety
specifications to existential reasoning over our incorrectness
pattern, it becomes easier to obtain a precise characterization of
exactly the unsafe behaviors one wish to detect.

\section{Trace-Augmented Refinement Types as Incorrectness Witnesses}

In this work, we propose a type-theoretic approach for searching
witnesses to this novel form of trace-based incorrectness
specifications.  Inspired by the recent work on Incorrectness Logic
(IL) \cite{ohearnIncorrectnessLogic2019}, our type system views types
through the lens of \emph{must}-style underapproximate reasoning,
ensuring that types serve as true witnesses to program behaviors.  One
example of \emph{must}-style types is Coverage Type
\cite{zhouCoveringAllBases2023} -- an underapproximate refinement type
denoting a set of values the associated term must be able to produce:
\begin{mathpar}
  \entails 1 : \tUnder{\type{int}\mid\nu=1} \and
  \notentails 1 : \tUnder{\type{int}}
\end{mathpar}
To track effectful behaviors in programs, we again leverage traces.  A
trace of \eff{put} events with their arguments properly qualified,
\eg, \cref{eq:bad-trace-leading}, simply underapproximates the effect
of a program that can perform these \eff{put} operations in sequence.
We additionally keeps track of the traces of events those operations
are conditioned upon before their invocations.  Consider a \eff{get}
operation, as long as the trace leading to the operation correctly
associates argument $k$ with some value denoted by a ghost variable
$v$ (\ie, admissible by \refLinkTo{k}{v}), its effect is exactly
captured by a singleton event $\inangle{v{\gets}\eff{get}~k}$:
\begin{equation*}
  \eff{get}:
  v{:}\type{addr}\darr k{:}\type{addr}\arr
  \triple{\refLinkTo{k}{v}}
  {\tUnder{\type{addr}\mid\nu=v}}
  {\inangle{v{\gets}\eff{get}~k}}
\end{equation*}
The elimination of such function types allows us to instantiate not
only the parameter but also the ghost variable and the context, thus
underapproximating the behavior specified.  For example, we may assume
that the last event is a corresponding $\eff{put}$ storing $1$ and
then the invocation of $\eff{get}$ on $1$ can be typed as:
% Types are then augmented with such traces to capture the
% underapproximation of both pure and effectful program behavior:
% \begin{mathpar}
%   \entails \eff{get}~1;\eff{get}~1 :
%   \triple{\refLinkTo{1}{1}}{\tUnder{\nu=1}}{
%     \inangle{1{\gets}\eff{get}~1} \inangle{1{\gets}\eff{get}~1}}
%   \end{mathpar}
\begin{equation*}
  \entails \eff{get}~1:
  \triple{\bullet^*\inangle{\eff{put}~1~1}}{\tUnder{\nu=1}}{
    \inangle{1{\gets}\eff{get}~1}}
\end{equation*}
Such types, when composed together, may serve as witnesses that the
program is incorrect provided that the traces within the types overlap
with the set of traces satisfying the incorrectness specification.
This is in contrast to trace-based verification techniques
\cite{skalkaHistoryEffectsVerification2004,nanjoFixpointLogicDependent2018,songAutomatedTemporalVerification2022,sekiyamaTemporalVerificationAnswerEffect2023}
where an overapproximation of effects is inferred and its inclusion in
the safety property is checked.  Underapproximate reasoning, unlike
overapproximate reasoning, is inherently non-deterministic; we may
need to blindly enumerate different underapproximate types before
finding the one that indeed witnesses the incorrectness.

\section{Algorithmic Incorrectness Witness Synthesis via Type
  Inference}

To reduce the inherent non-determinism in underapproximate reasoning,
we develop a type inference algorithm that harnesses the incorrectness
specification and derives only types that concern the incorrectness.
Consider the following program that manipulates a linked list via
operations to the key-value store $\eff{Nxt}$:
\begin{equation}\label{incorrect-program}
  \letin{n_1}{\eff{get}~n_0}
  \letin{n_2}{\eff{get}~n_1}
  \eff{put}~n_0~n_2
  \tag{$e_\textsf{bad}$}
\end{equation}
The incorrectness specification \ref{not-unique} can be naturally
decomposed into \ref{link-a-to-b} followed by \ref{bad-link-to-b},
defined as:
\newcommand{\rgxBadLink}[2]{\EM{\FA_{#2 \curvearrowleft \cancel{#1}}}}
\newcommand{\refBadLink}[2]{\hyperref[bad-link-to-b]{\rgxBadLink{#1}{#2}}}
\begin{equation}\label{bad-link-to-b}
  (\lneg\inangle{\eff{put}~a~\_})^*
  \inangle{\eff{put}~\cancel{a}~b}\ \bullet^*
  \tag{$\rgxBadLink{a}{b}$}
\end{equation}
Here we use regexes for simplicity while our inference algorithm
operates over their SRE representations of regexes whose edges are
labeled by qualified events as in regexes; any state in the SFA
presenting the incorrect traces splits the automaton into halves.
Before type inference, we hypothesize that that some trace admissible
by \ref{link-a-to-b} allows the execution of \ref{incorrect-program}
to produce some other trace admissible by \ref{bad-link-to-b}; such
two traces suffice to witness the incorrectness.  The decomposition of
the incorrectness specification allows us to underapproximate the
behavior of \eff{get}s with respect to the incorrectness.

The hypothesized context enables intelligent underapproximation of
traces that may reach the program.  For example, the first \eff{get}'s
behavior is known under the context \refLinkTo{n_0}{n_1} where we
simply let $n_1$ be the ghost variable denoting the stored value.  As
our hypothesis is only interested in the program's behavior under the
context \ref{link-a-to-b}, we underapproximate the \eff{get}'s
behavior by finding the overlap between \ref{link-a-to-b} and
\refLinkTo{n_0}{n_1} via the canonical automata intersection
algorithm.
\begin{multline*}
  a{:}\tUnder{\top_\type{addr}},b{:}\tUnder{\nu\neq a},
  n_0{:}\tUnder{\top_\type{addr}},n_1{:}\tUnder{\top_\type{addr}} \entails \\
  \eff{get}~n_0:
  \triple{%
    \refLinkTo{a}{b}\wedge\refLinkTo{n_0}{n_1}%
  }{\tUnder{\nu=n_1}}{%
    \inangle{n_1{\gets}\eff{get}~n_0}%
  }
\end{multline*}
The traces leading to the second \eff{get} are then captured by a new
context
$(\refLinkTo{a}{b}\wedge\refLinkTo{n_0}{n_1})
\inangle{n_1{\gets}\eff{get}~n_0}$.  We again underapproximate the
second \eff{get}'s behavior under context \refLinkTo{n_1}{n_2} with
respect to the new context.  Instead of simply performing automata
intersection again, we leverage the qualified events that label the
edges and abduce, as we relate edges from two SFAs, that $n_1$ ($n_2$)
denotes the same node as $a$ ($b$).
\begin{multline*}
  a{:}\tUnder{\top_\type{addr}},b{:}\tUnder{\nu\neq a},
  n_0{:}\tUnder{\top_\type{addr}}, n_1{:}\tUnder{\nu=a}, n_2{:}\tUnder{\nu=b}
  \entails \\ \eff{get}~n_1: \\
  \triple{%
    (\refLinkTo{a}{b}\wedge\refLinkTo{n_0}{n_1})
    \inangle{n_1{\gets}\eff{get}~n_0}%
  }{\tUnder{\nu=n_2}}{%
    \inangle{n_2{\gets}\eff{get}~n_1}%
  }
\end{multline*}

Moreover, the hypothesized effect enables intelligent
underapproximation of the effect produced by the program.  The pattern
$(\lneg\inangle{\eff{put}~a~\_})^*$ in the hypothesized effect
\ref{bad-link-to-b} specifies that it is safe to ignore events that do
not intervene with the existing link between $a$ and $b$, including
the effect of the two \eff{get} operations.  Then we just need to
match the last operation $\inangle{\eff{put}~n_0~n_2}$ against
$\inangle{\eff{put}~\cancel{a}~b}$ to show that the program is
incorrect.  Again by abduction, we hypothesize that $n_0$ is some node other than $a$.
\begin{multline*}
  a{:}\tUnder{\top_\type{addr}},b{:}\tUnder{\nu\neq a},
  n_0{:}\tUnder{\nu\neq a}, n_1{:}\tUnder{\nu=a}, n_2{:}\tUnder{\nu=b}
  \entails \\ \eff{put}~n_0~n_2: 
  \triple{%
    (\refLinkTo{a}{b}\wedge\refLinkTo{n_0}{n_1})\\
    \inangle{n_1{\gets}\eff{get}~n_0}
    \inangle{n_2{\gets}\eff{get}~n_1} 
  }{\tUnder{\type{unit}}}{%
    \inangle{\eff{put}~n_0~n_2}%
  }
\end{multline*}
As we return to the top level, we eliminate local variables from the
typing context:
\begin{multline*}
  a{:}\tUnder{\top_\type{addr}},b{:}\tUnder{\nu\neq a},
  n_0{:}\tUnder{\nu\neq a} \entails\ $\ref{incorrect-program}$\ : \\
  \triple{%
    (\refLinkTo{a}{b}\wedge\refLinkTo{n_0}{a})
    \inangle{a{\gets}\eff{get}~n_0}
    \inangle{b{\gets}\eff{get}~a} 
  }{\tUnder{\type{unit}}}{%
    \inangle{\eff{put}~n_0~b}%
  }
\end{multline*}
The inferred type describes a set of incorrect traces that not only
represent true behavior of \ref{incorrect-program} but also are
admissible by the incorrectness specification \ref{not-unique}.
Intuitively, our type inference algorithm facilitates a trace-guided
compositional reasoning of both program behaviors and the
incorrectness with the purpose of finding their overlap sooner.

% Focusing on incorrectness and falsification provides us a fundamental
% advantage over the verification techniques that has a deep connection
% with the existential reasoning of incorrectness -- it suffices to find
% one incorrect trace that the program can produce.

\section{Conclusion}

We advocate the adoption of incorrectness reasoning principal for
trace-based specification as it enables existential and local
reasoning of patterns in regex-like specification language.  We
further propose a type inference algorithm that automatically
decomposes this novel form of trace-based incorrectness specification
and derive composable type-based witnesses to the incorrectness from
subprograms.  As a result, the types inferred from programs correctly
witness their incorrect behavior.